# QoS-Aware Radio Access Technology (RAT) Selection in Hybrid Vehicular Networks

Zeeshan Hameed Mir[1], Jamal Toutouh[2], Fethi Filali[1], and Enrique Alba[2]

[1]Qatar Mobility Innovations Center (QMIC), Qatar Science and
Technology Park (QSTP), PO Box 210531, Doha, Qatar
{zeeshanh,filali}@qmic.com
[2]Dept. de Lenguajes y Ciencias de la Computación,
University of Malaga, Malaga, Spain
{jamal,eat}@lcc.uma.es

# QoS-Aware Radio Access Technology (RAT) Selection in Hybrid Vehicular Networks

Zeeshan Hameed Mir[1], Jamal Toutouh[2], Fethi Filali[1], and Enrique Alba[2]

[1] Qatar Mobility Innovations Center (QMIC), Qatar Science and
Technology Park (QSTP), PO Box 210531, Doha, Qatar
{zeeshanh,filali}@qmic.com
[2] Dept. de Lenguajes y Ciencias de la Computación,
University of Malaga, Malaga, Spain
{jamal,eat}@lcc.uma.es

**Abstract.** The increasing number of wireless communication technologies and standards bring immense opportunities and challenges to provide seamless connectivity in Hybrid Vehicular Networks (HVNs). HVNs could not only enhance existing applications but could also spur an array of new services. However, due to sheer number of use cases and applications with diverse and stringent QoS performance requirements it is very critical to efficiently decide on which radio access technology (RAT) to select. In this paper a QoS-aware RAT selection algorithm is proposed for HVN. The proposed algorithm switches between IEEE 802.11p based ad hoc network and LTE cellular network by considering network load and application's QoS requirements. The simulation-based studies show that the proposed RAT selection mechanism results in lower number of Vertical Handovers (VHOs) and significant performance improvements in terms of packet delivery ratio, latency and application-level throughput.

## 1 Introduction

Recently, several studies [1–4] urged to combine different radio access technologies into a unified Hybrid Vehicular Networking (HVN) architecture. Multiple radio access technologies for vehicular communications could not only enhance several existing applications in the road safety, traffic efficiency, and infotainment domains, but could also spur an array of new services. However, successful implementation of such hybrid architecture is often attributed to efficient management of combined radio resources. Central to this are the Radio Access Technology (RAT) selection algorithms which carefully choose the most suitable access technology while preserving the connectivity through Vertical Handover (VHO). For this purpose, standards like Communications Architecture for Land Mobile environment (CALM) [5] and Media Independent Handover (MIH) or IEEE 802.21[6] can be utilized in the vehicular networking environment. It is highly desirable to limit the number of unnecessary VHOs because of the higher VHO signaling cost and the delay incurred which often translate into data throughput lost.

To fully exploit the potential of HVN architecture we consider to combine IEEE 802.11p [7] based ad hoc network and infrastructure-based LTE [8] cellular network. It is envisioned that the proposed HVN architecture would support different types of uses cases and applications such as road safety, traffic efficiency and infotainment. The applications exchange messages called *beacons* at regular intervals and the frequency with which beacons are transmitted is termed as *beaconing frequency*. Since each application has its own set of functional and performance requirements, typically the beaconing frequency varies between 1Hz to 10Hz. The IEEE 802.11p standard provides ad hoc networking capabilities to exchange messages directly and is particularly suitable for low to medium range and delay sensitive vehicular networking applications. On the other hand LTE by 3GPP promises to offer medium to large communication range, higher data rates with moderate delays. Each beacon is sent to the base station (eNodeB), which traverse through the core network before reaching an Intelligent Transportation Systems (ITS) Server. The ITS Server acts as a rendezvous between the senders and receivers and transfers the beacon messages back to sender's neighboring vehicles through broadcasting by the eNodeB.

RAT selection algorithms have been extensively studied in heterogeneous mobile networks [9]. In context of vehicular communications most of studies focused on protocol switching algorithms between co-located WLAN and WWAN access technologies. To provide seamless connectivity, the VHO decisions are made either by predicting location information or by measuring multiple performance criteria. The former approach relies excessively on the accuracy of prior knowledge such as call duration and hot-spot dwell time. Whereas the later approach results in higher VHO frequency [10].

In this paper we proposed an efficient QoS-aware RAT selection algorithm which switches between IEEE 802.11p and LTE access technologies by performing VHO. Coordinated by a Distributed Radio Resource Management (DRRM) [9] entity, the IEEE 802.11p interface exchanges periodic beacon messages directly in an ad hoc manner. Alternatively, based on the network load and QoS requirements imposed by vehicular networking applications, the DRRM can choose to perform vertical handover to the LTE interface. However, in order to lower the frequency of VHO and associated cost the proposed mechanism first tries to locally solve the network load issue through beaconing frequency reduction [11] as permissible by application's QoS requirements. The study shows that the *QoS-aware RAT selection* and *Beaconing Frequency Adaptation* mechanism result in fewer number of VHOs while satisfying QoS requirements by different types of applications. The proposed mechanism preformed significantly well in terms of packet delivery ratio, latency and application-level throughput.

The rest of this paper is organized as follows. The related work is described in Section II and the proposed QoS-aware RAT selection algorithm is explained in Section III. The performance evaluation is provided in Section IV. Finally, the paper is concluded in Section V.

## 2   Related Work

Much work has been done in heterogeneous mobile networks, and efficient selection of radio access technology (RAT) is one of most addressed research issues. This section summarizes the work on RAT selection in mobile networks and reviews studies on heterogeneous wireless network in vehicular environment.

Studies on RAT selection in mobile networks can be largely classified into three categories: i.e., random, single-criterion, and multiple-criteria based algorithms [9]. In random RAT selection, all new or VHO (vertical handover) calls are handled by one of the available access technology. The simplicity of implementation comes at the cost of higher call blocking and dropping probability. The single and multiple criteria approaches differ in using either only one or several requirements while evaluating the suitability of a network. Work falling into the former category takes into account a single criterion such as load balancing, coverage, user satisfaction and service types. The later approaches consider diverse performance metrics and parameters to decide on the candidate access technology. It includes Analysis Hierarchy Process (AHP), Simple Additive Weighting (SAW), Multiplicative Exponent Weighting (MEW) and fuzzy logic-based methods. The single criterion approaches improve system performance in certain specific aspects while multiple criteria approaches provide optimized solutions, however they are complicated to implement.

Heterogeneous network architectures for vehicular communications have been realized by mean of accessing Wireless Local Area Network (WLAN) and Wireless Wide Area Network (WWAN) technologies in parallel. Earlier work in this domain exploit co-located Access Points (APs) and cellular technologies such as CDMA2000 [12] and UMTS [13] to implement hybrid communication paradigm for vehicular networking. These studies were focused on providing hybrid V2X communication protocol with the help of VHO between Vehicle-to-Vehicle (V2V) and Vehicle-to-Infrastructure (V2I) communications. However the main difference is the VHO strategy which they employed. Anna et. al. in [10] proposed a switching protocol decision metric based on the cost function which is measured by taking into account delay and radio resource utilization time associated with the alternative paths. The VHO decision based on multiple performance criteria results in higher number of VHOs (i.e., ping-pong effect) which often leads to protocol instability [13].

Similarly, the VHO decision in [12] is based on metrics like service type, congestion and location prediction. Based on the predicted dwell time at a hot-spot, if the user is expected to stay, WLAN become the preferred choice; otherwise WWAN is the preferred choice. For real-time type services, WWAN is selected and for non-real-time service WLAN, only if WWAN is not congested. In Hasib's work it is assumed that the call duration and the hot-spot dwell time are known in advance, which is quite hard to achieve because vehicular network topology changes rapidly and in an unpredictable manner.

# 3 Proposed RAT Selection Algorithm

It is assumed that each end user implements a distributed strategy for managing multiple radio resources, called *Distributed Radio Resource Management (DRRM)* entity [9]. Unlike the *centralized* or *hierarchical* approaches, the DRRM entity independently makes the RAT selection decisions. DRRM is not only responsible for providing coordination among participating access technologies which are managed locally by their respective *Radio Resource Management* (RRM) entities but also among different DRRM entities.

## 3.1 QoS-Aware RAT Selection

The IEEE 802.11p RRM entity is in charge of analyzing the IEEE 802.11p network load by using the *Network Load Monitor* (NLM) mechanism, which monitors the length of the queues in order to determine the current network load. If the queue lengths are below a given threshold limit, the DRRM entity broadcasts beacons via its IEEE 802.11p interface. However, if the queue lengths exceed certain threshold limit, it considers that the current network load could lead to a congestion situation which in turns causes severe performance degradation. The threshold limit is defined by the parameter *NLM-Threshold* which often varies between 80% to 90% of the total queue capacity.

To locally resolve the network load problem, the DRRM entity initiates the Beaconing Frequency Adaptation (BFA) mechanism which allows the applications to adapt and operate the IEEE 802.11p interface with new set of communication requirements. Therefore, the use of LTE radio access technology and the associated cost of performing VHOs can be minimized. The proposed QoS-aware RAT selection algorithm can be summarized as follow:

**Step 1:** On finding that the network load exceeds the predefined threshold limit i.e., NLM-Threshold, the DRRM entity triggers the BFA mechanism which reduces the number of beacons transmitted in the ad hoc network. The BFA mechanism adjusts the beaconing frequency of an application by adapting its QoS requirements.

**Step 2:** If the local adjustment of an application's beaconing frequency fails to lower the network load enough, in the next step the DRRM entity initiates a request to the neighboring vehicles to apply the BFA mechanism. On receiving such requests the DRRM entity at each neighboring vehicle tries to adapt its application's beaconing frequency.

**Step 3:** If applying above two steps brings down the network load below the NLM-Threshold, the DRRM entity suggests to broadcast the beacons through its IEEE 802.11p interface with the reduced beaconing frequency. However, if the network load is still high and the beaconing frequency cannot be further reduced without significantly sacrificing the QoS requirements, a VHO is performed. All the subsequent beacons are sent through the LTE interface for certain duration. This duration is usually short mainly due to vehicular mobility where topology and the network load changes rapidly. Fig. 1, illustrates the QoS-aware RAT selection algorithm.

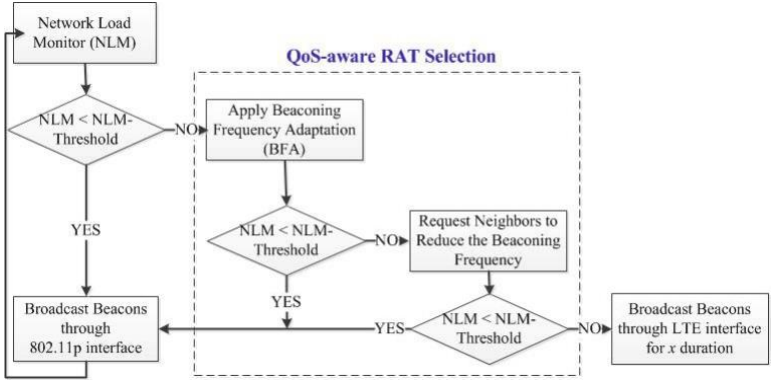

**Fig. 1.** Flowchart: QoS-aware RAT selection algorithm

### 3.2 Beaconing Frequency Adaptation (BFA)

By adjusting the QoS requirements of an application according to the network load conditions the Beaconing Frequency Adaptation (BFA) mechanism improves the network efficiency and reduces the number of VHOs and the associated cost. Fig. 2 summarizes the BFA procedure which is either triggered locally or as a result of a request received by one or more of the neighboring vehicles. It starts with the applications reporting their current QoS requirements to the DRRM entity. It is assumed that these requirements can be characterized using three parameters namely initial beaconing frequency *bFreqInitial*, beaconing frequency reduction factor *rFactor* and maximum beaconing frequency reduction tolerance *rTolerance*. Based on these parameters, the BFA mechanism calculates the reduced beaconing frequency *bFreqReduced* given as,

$$bFreqReduced = [bFreqReduced - (rFactor\% \times bFreqInitial)]$$

After initially setting bFreqReduced to bFreqInitial, an iterative procedure is carried out by the DRRM entity which checks whether the reduction in initial frequency is permissible as per application's QoS requirements defined by the parameter rTolerance. If the beaconing frequency can be reduced, the DRRM entity requests the application layer to decrease its initial beaconing frequency by the given rFactor. The BFA mechanism finishes either when an application can't tolerate further decrease in its beaconing frequency or the network load is reduced to the level where beacons can be efficiently transmitted over the IEEE 802.11p interface.

An application's QoS requirements are preserved by using rTolerance and rFactor parameters. The parameter rTolerance is the maximum tolerable reduction in beaconing frequency by an application, given as the percentage of the initial beaconing frequency (bFreqInitial). This parameter ensures that the application functions properly along with its QoS requirements without losing the useful information to the neighboring vehicles. Whereas the parameter rFactor is used to gradually decrease the beaconing frequency in relatively smaller steps instead of reducing it directly by the rTolerance. For example, in a scenario where an application sets bFreqInitial at 10Hz, with rTolerance and rFactor equal to 50% and 25% of the bFreqInitial, respectively. For the first, second and third iteration of the BFA mechanism the parameter bFreqReduced is calculated as 8Hz, 6Hz, and 4 Hz, respectively. From the implementation perspective two timers were defined as well.

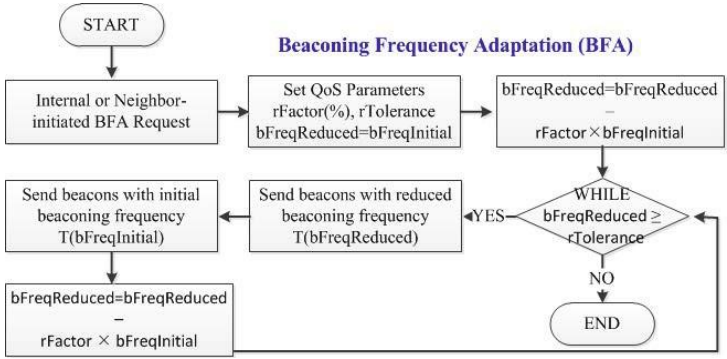

**Fig. 2.** Flowchart: Beaconing Frequency Adaptation (BFA)

1. The timer T(bFreqReduced) defines the maximum time period an application can operate at reduced beaconing frequency without compromising the QoS requirements. During this duration the vehicles would broadcast beacons with a frequency lower than the value initially set, as illustrated in Fig. 3(a).
2. The timer T(bFreqInitial) defines the minimum time period that an application must operate at initial beaconing frequency right after the T(bFreqReduced) epoch, as shown in Fig. 3(b).

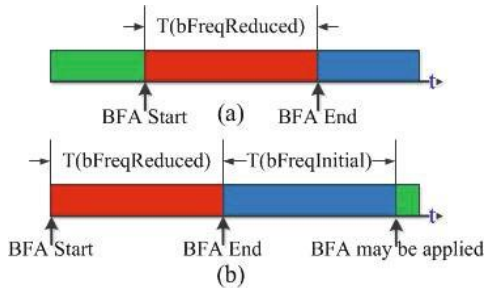

**Fig. 3.** The timers: (a) T(bFreqReduced) and (b) T(bFreqInitial)

## 4  Performance Evaluation

This section evaluates and compares the QoS-aware RAT selection algorithm using self-developed simulation tool and consists of two types of simulation studies.

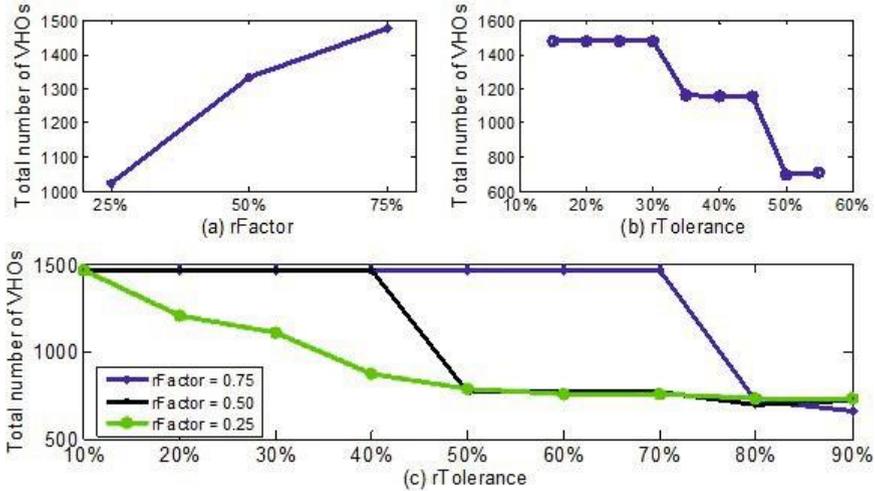

**Fig. 4.** Impact of QoS parameters: (a) rFactor, (b) rTolerance and (c) Combined rFactor and rTolerance vs. Number of VHOs

Firstly, the proposed algorithm is evaluated for different values of QoS parameters such as rFactor, rTolerance, T(bFreqReduced) and T(bFreqInitial) and their impact on the number of VHOs. Secondly, a comparative study is provided which evaluated several RAT selection mechanisms in terms of packet delivery ratio (PDR), latency and application-level throughput or goodput.

### 4.1 Simulation Environment

The road network represents a highway scenario of length 1km with three lanes each 5m wide. There are 150 vehicles participating, with varying speed between minimum 50km/h to maximum 130km/h. For modeling IEEE 802.11p simulation the Three-Log Distance propagation model is used with 5.8GHz radio operating at 6Mbps data rate. The communication range is set to 250m. As for the of LTE part a simplified Radio Access Network (RAN) is modeled with one eNodeB, thus a single cell environment operating at 900 MHz with a bandwidth of 10Mhz. The application running at each vehicle transmits 100 bytes beacons at varying beaconing frequencies. Each simulation runs for 100 seconds, and the obtained results are the averaged over 10 different simulation instances.

### 4.2 Simulation Study

**Impact of QoS Parameters.** Fig. 4(a) shows that with the increase in rFactor, the number of VHO increases. Even with a moderate reduction factor of 25% the number of VHOs are quite high. For more aggressive increments in rfactor the number of VHOs grows significantly. Since the maximum beaconing frequency is set to 10Hz, the rTolerance is achieved at a very early iterations of the BFA mechanism. Therefore for the subsequent iterations, the parameter rTolerance wouldn't let BFA mechanism to apply. This causes network load to increase and thus high number of VHOs. Fig. 4(b) shows that overall with the increase in rTolerance, the number of VHOs decreases. The higher rTolerance values let applications to reduce the beaconing frequency at maximum which leads to lower network load and therefore less number of VHOs. Fig. 4(c) shows the combined effect of parameters rTolerance and rFactor on the number of VHOs. The constant number of VHOs is due to smaller changes between the consecutive values of rTolerance. Unless the rTolerance parameter value is considerably greater than that of rFactor (i.e., applications are more flexible) the effect of increasing reduction factor on the number of VHOs remains less significant.

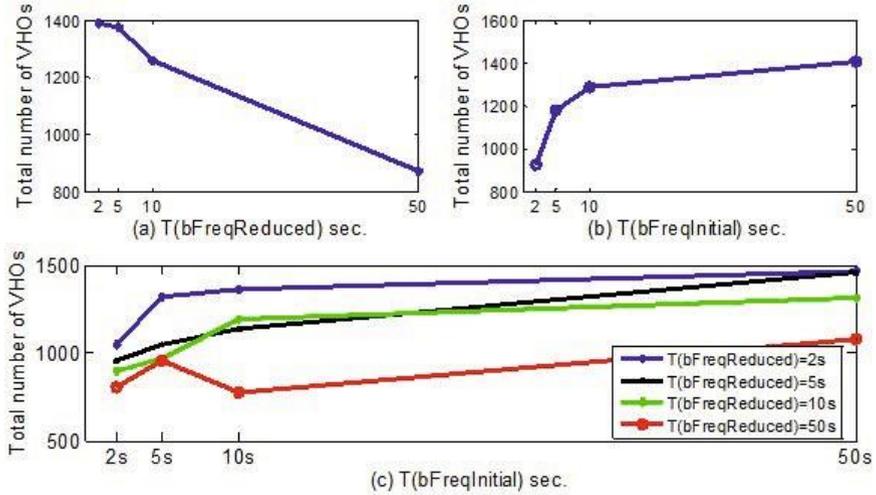

**Fig. 5.** Impact of timer values: (a) T(bFreqReduced), (b) T(bFreqInitial) and (c) Combined T(bFreqReduced) and T(bFreqInitial) vs. Number of VHOs

Fig. 5(a) shows that the increase in T(bFreqReduced) (i.e., an application can operate at reduced beaconing frequency for longer duration without sacrificing the QoS requirements) results in lower number of VHOs. Since, an application can tolerate to operate at reduced beaconing frequency longer, there are fewer chances of network load to increase beyond the threshold limits, therefore requires fewer number of VHOs. Fig. 5(b) shows that the increase in T(bFreqInitial) (i.e., an application is required to stay longer at the initial beaconing frequency right after T(bFreqReduced) duration) causes higher number of VHOs. An application tends to operate at original beaconing frequency for longer duration, which cause significant increase in network load. The frequent violation of the threshold limit which is often can't be handled by the BFA procedure result in higher number of VHOs. Generally, as the T(bFreqInitial) increases, for all the simulated values of T(bFreqReduced) the number of VHOs increases, as given in Fig. 5(c). No matter for what duration of time an application can tolerate the reduction of original beaconing frequency, if it requires attaining the initial beaconing frequency longer afterwards this would increase the network load to the level where VHO cannot be avoided. Similarly, as the timer T(bFreqReduced) value increases, for all the simulated values of timer T(bFreqInitial), the number of VHO decreases. Shorter an application can tolerate to stay at original beaconing frequency, fewer the chance are for the network load to increase beyond the specified threshold limit. Therefore fewer numbers of VHOs are required to satisfy the application QoS requirements.

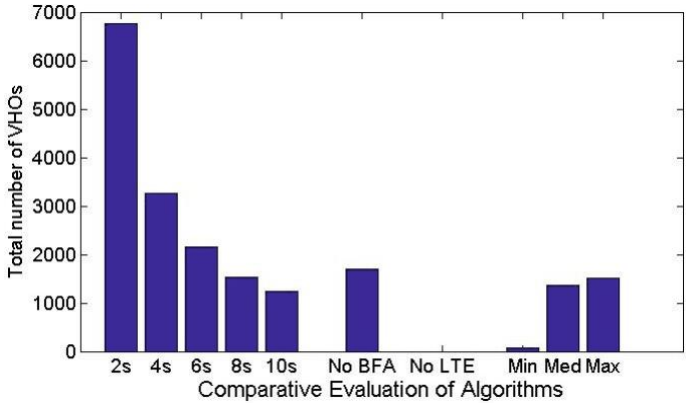

**Fig. 6.** Number of VHO for different RAT selection schemes

**Comparative Study.** For a comparative study, the proposed QoS-aware RAT selection mechanism is compared with number of other schemes, including:

1. Periodic RAT selection: The decision of switching between two access technologies is carried out in discrete period in time (i.e., proactive handover [14]). Simulations were performed for several epochs from 2s to 10s.
2. No BFA: In this scheme, every time the network load exceeds the predefined NLM-Threshold value, the algorithm performs VHO. This is similar to load balancing among multiple RAT with a fixed threshold value [15].
3. No LTE: It is assumed that there is no LTE interface available i.e., only the performance of IEEE 802.11p based vehicular ad hoc network is evaluated.

Fig. 6 shows the number of VHOs performed by each scheme. As expected, for the periodic RAT selection this metric value decreases with the increase in periodicity interval. The number of VHOs in the No BFA scheme is quite comparable with periodic scheme for values between 6s and 8s. For the proposed QoS-aware RAT selection mechanism the values of minimum, median, and maximum number of VHOs are plotted. The minimum number of VHOs is significantly lower than any other algorithm, whereas the median and maximum values are equivalent to the periodic scheme with period in between 8s and 10s. The fewer number of VHO signifies less reliance on the LTE interface while reducing VHO cost in terms of latency and the corresponding lost in data throughput.

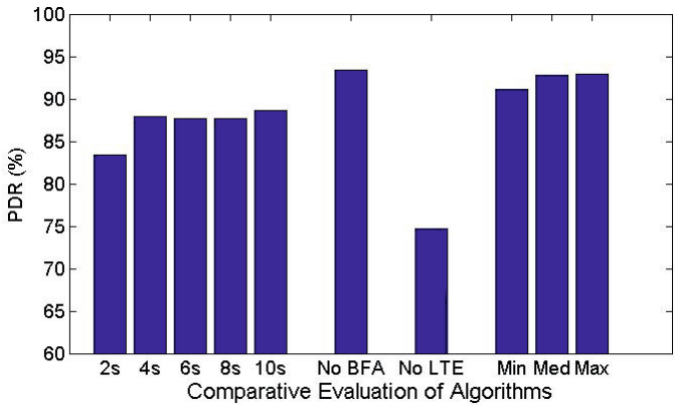

**Fig. 7.** Packet Delivery Ratio (%) for different RAT selection schemes

The number of VHOs, significantly impact other performance metrics. Fig. 7 illustrates the packet delivery ratio (PDR) statistics. The PDR for the periodic switching scheme varies between 80% to 90%. Higher number of VHOs results in higher delays which in turn result in lower delivery of beacons that could have been delivered during the VHO delays. The PDR for No BFA and our proposed mechanism are quite comparable whereas No LTE suffers from severe reliability issues which are mainly due to higher network load.

As shown in Fig. 8, in No BFA the performance in terms of lower number of VHOs and higher PDR is completely offset due to higher latency. No LTE, results in lower beacon exchange latency because of the direct communication among the vehicles. By switching access technologies only when it is necessary without sacrificing the QoS requirements for an application, overall the proposed RAT selection mechanism achieved considerably lower latency.

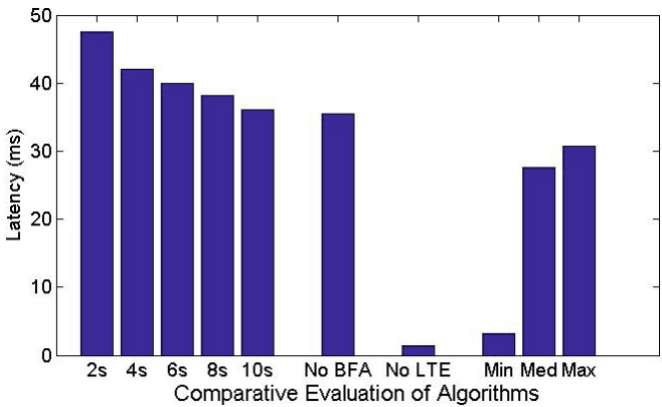

**Fig. 8.** Latency for different RAT selection schemes

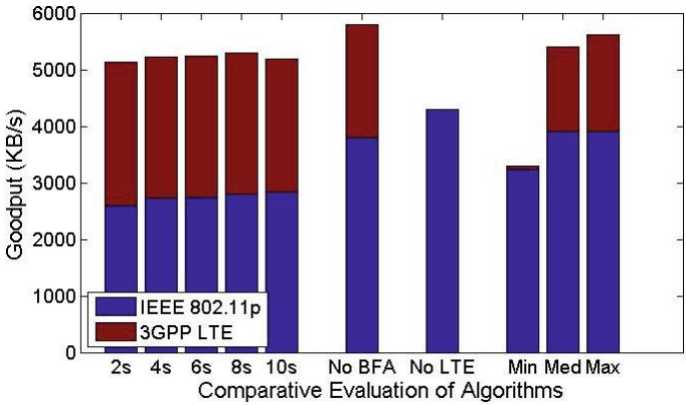

**Fig. 9.** Goodput for different RAT selection schemes

Fig. 9 compares different schemes in terms of goodput. The bar plots are stacked comparing the contribution of each access technology to the total achievable goodput. Schemes with dual-interfaces attain comparable goodput, however the contribution by each access technology differs. In periodic switching role of each interface is similar. No BFA performs well, however at the cost of significantly higher delays with higher involvement of the LTE interface. For the proposed RAT selection mechanism the IEEE 802.11p interface delivers most of beacons whereas the use of LTE interface is reasonably lower than all other schemes.

## 5  Conclusion

In hybrid vehicular networks, efficient RAT selection while simultaneously achieving QoS requirements is a challenging task. In this paper a RAT selection algorithm is proposed in context of IEEE 802.11p based infrastructure-less and LTE based infrastructure-based cellular networks. The key idea is to use parameters like network load and desired QoS application requirements to make switching decisions between the two radio access technologies. In order to reduce the number of vertical handovers (VHOs), the algorithm reduces the beaconing frequency as permissible by the application's QoS requirements. Simulations show that by locally resolving the network load issues result in lower VHO frequency. Moreover, a comparative study with several other competitor mechanisms show that the proposed work achieve considerable performance improvements in terms of packet delivery ratio, latency and application-level throughput or goodput.

**Acknowledgments.** This work was made possible by NPRP Grant No.: 5-10801-186 from the Qatar National Research Fund (a member of The Qatar Foundation). The statements made herein are solely the responsibility of the authors.

J. Toutouh was supported by Grant AP2010-3108 and Est13/00988 of the Spanish Ministry of Education. This research has been partially funded by project number 8.06/5.47.4142 in collaboration with the VSB-Technical University of Ostrava) and UMA/FEDER FC14-TIC36. University of Malaga, International Campus of Excellence Andalucia Tech.